\documentclass[twocolumn,preprintnumbers,amsmath,amssymb]{revtex4}

\usepackage{graphicx}
\usepackage{dcolumn}
\usepackage{bm}

\begin{document}

\title{Proton distribution radii of $^{16-24}$O : signatures of new shell closures and neutron skin} 

\author {S. Kaur$^{1,2}$, R. Kanungo$^{1,3}$, W. Horiuchi$^{4,5,6}$, G. Hagen$^{7,8,3}$, J. D. Holt$^3$, B. S. Hu$^3$, T. Miyagi$^{9,10}$, T. Suzuki$^{11}$, F. Ameil$^{12}$, J. Atkinson$^1$, Y. Ayyad$^{13}$, S. Bagchi$^{1,12}$, D. Cortina-Gil$^{13}$, I. Dillmann$^{12,14}$, A. Estrad\'e$^{1,12}$,  A. Evdokimov$^{12}$, F. Farinon$^{12}$, H. Geissel$^{12,14}$, G. Guastalla$^{12}$,  R. Janik$^{15}$, R. Kn\"obel$^{12}$, J. Kurcewicz$^{12}$, Yu. A. Litvinov$^{12}$, M. Marta$^{12}$, M. Mostazo$^{13}$, I. Mukha$^{12}$, C. Nociforo$^{12}$, H.J. Ong$^{16}$, T. Otsuka$^{17,18}$, S. Pietri$^{12}$, A. Prochazka$^{12}$,  C. Scheidenberger$^{12,14}$, B. Sitar$^{15}$, P. Strmen$^{15}$*,  M. Takechi$^{12}$,  J. Tanaka$^{16}$,  I. Tanihata$^{16,19}$, S. Terashima$^{19}$, J. Vargas$^{13}$, H. Weick$^{12}$,  J. S. Winfield$^{12}$*}
\thanks{Deceased}

\affiliation{$^1$Astronomy and Physics Department, Saint Mary's University, Halifax, NS B3H 3C3, Canada}
\affiliation{$^{2}$Department of Physics and Atmospheric Science, Dalhousie University, Halifax, NS B3H 4R2, Canada}
\affiliation{$^3$TRIUMF, Vancouver, BC V6T 4A3, Canada}
\affiliation{$^4$Department of Physics, Osaka Metropolitan University, Osaka 558-8585, Japan}
\affiliation{$^5$ Nambu Yoichiro Institute of Theoretical and Experimental Physics (NITEP),
                 Osaka Metropolitan University, Osaka 558-8585, Japan}
\affiliation{$^6$ Department of Physics, Hokkaido University, Sapporo 060-0810, Japan}
\affiliation{$^7$Physics Division, Oak Ridge National Laboratory, Oak Ridge, TN 37831, USA }
\affiliation{$^8$Department of Physics and Astronomy, University of Tennessee, Knoxville, TN 37996, USA}
\affiliation{$^9$Technische Universit\"at Darmstadt, Department of Physics, 64289 Darmstadt, Germany}
\affiliation{$^{10}$ExtreMe Matter Institute EMMI, GSI Helmholtzzentrum f\"ur Schwerionenforschung GmbH, 64291 Darmstadt, Germany}
\affiliation{$^{11}$Department of Physics, Nihon University, Setagaya-ku, Tokyo 156-8550, Japan}
\affiliation{$^{12}$GSI Helmholtzzentrum f\"ur Schwerionenforschung, D-64291 Darmstadt, Germany}
\affiliation{$^{13}$Universidad de Santiago de Compostela, E-15706 Santiago de Compostella, Spain}
\affiliation{$^{14}$Justus-Liebig University,  35392 Giessen, Germany}
\affiliation{$^{15}$Faculty of Mathematics and Physics, Comenius University, 84215 Bratislava, Slovakia}
\affiliation{$^{16}$RCNP, Osaka University, Mihogaoka, Ibaraki, Osaka 567 0047, Japan}
\affiliation{$^{17}$Department of Physics, University of Tokyo, Bunkyo-ku, Tokyo 113-0033, Japan}
\affiliation{$^{18}$RIKEN Nishina Center, Hirosawa, Wako, Saitama 351-0198, Japan}
\affiliation{$^{19}$School of Physics and Nuclear Energy Engineering and IRCNPC, Beihang University, Beijing 100191, China}

\date{\today}

\begin{abstract}
  The root mean square radii of the proton density distribution in $^{16-24}$O derived from measurements of charge changing cross sections with a carbon target at $\sim$900$A$ MeV together with the matter radii portray thick neutron skin for $^{22 - 24}$O despite $^{22,24}$O being doubly magic. Imprints of the shell closures at $N$ = 14 and 16 are reflected in local minima of their proton radii that provide evidence for the tensor interaction causing them. The radii agree with {\it ab initio} calculations employing the chiral NNLO$_{sat}$ interaction, though skin thickness predictions are challenged. Shell model predictions agree well with the data.
 \end{abstract}

\maketitle

Nuclear shell structure has profound impact in shaping the elemental abundance in the universe. Nuclei with filled proton and/or neutron shells, i.e. magic numbers,  play a significant role.  Doubly magic nuclei are key benchmarks for constraining the nuclear force and nuclear models. Oxygen isotopes have closed proton shell ($Z$ = 8). The doubly magic nature of $^{16}$O leads to its copious abundance hence enabling sustaining life in the universe. The rare isotopes are unveiling new nuclear shells and exotic neutron skin and halo structures.  At the edge of neutron binding, the neutron drip-line, a new magic number has surfaced at $N$ = 16 making the heaviest oxygen isotope $^{24}$O an unexpected doubly magic nucleus. A sub-shell closure at $N$ = 14 also emerges in $^{22}$O. Do these neutron shell closures impact the proton distribution? Does the doubly-magic nature of $^{22,24}$O hinder neutron skin formation? 

This Letter addresses the questions above through experimental determination of the root mean square radii of the point proton density distributions, henceforth referred to as point proton radii,  in $^{16,18-24}$O.  

The signature of shell closures $N$ = 50 and 82 is seen from a local dip in the proton radius for isotopes \cite{Angeli}. In neutron-rich light nuclei a new sub-shell gap at $N$ = 6 shows prominent minimum in the proton radii for He to B isotopes \cite{Angeli}. The proton radii of nitrogen isotopes hinted a dip at $N$ = 14 \cite{Bagchi2019}. If the possible origin of this sub-shell closure is due to the attractive isospin ($T$) = 0 $p-n$ tensor interaction it would be reflected also in the proton radii of neutron-rich oxygen isotopes. 

The new shell closure at $N$ = 16 is seen in the high excitation energy of the first excited state \cite{Hofman2009} of $^{24}$O and from the large 2$s_{1/2}$ orbital \cite{Kanungo2009} occupancy of the valence neutrons, reflected in the neutron removal momentum distribution. Proton inelastic scattering of $^{24}$O shows a small quadrupole deformation of 0.15(4) confirming a spherical shell closure at $N$ = 16 \cite{Tshoo2012}.
 
A sub-shell closure at $N$ = 14 for $^{22}$O is discussed from high energy of its 2$^+$ first excited state  \cite{Stanoiu2004} and a small quadrupole deformation parameter 0.26(4) \cite{Becheva2006} compared to $^{20}$O. Quasifree $(p,pn)$ neutron knockout \cite{DiazFernandez2018}  and neutron removal with carbon target \cite{Sauvan2004} from $^{22}$O result in a wider momentum distribution reflecting knockout of 1$d_{5/2}$ neutrons, consistent with $N$ = 14 sub-shell gap. A narrower momentum distribution  for $^{21}$N suggests reduction of $N$ = 14 shell gap in nitrogen. The quenching is derived from unbound states in $^{22}$N \cite{Strongman2009}. It is predicted that a 2$s_{1/2}$ - 1$d_{5/2}$ level inversion may occur in $^{20}$C. Proton knockout via $(p,2p)$ reactions show a larger cross section for $^{22,23}$O than $^{21}$N \cite{DiazFernandez2018} interpreted being due to more protons in the 1$p_{1/2}$ orbital in oxygen isotopes. The wider proton removal momentum distribution for $^{22}$O is qualitatively suggested to be due to its compact nature from filled valence shell for protons. However, that for $^{23}$O is indicated to be narrow, which remains to be understood. 

The large matter radii for $^{23}$O \cite{Kanungo2011} and $^{24}$O \cite{Ozawa2000,Ozawa2001} from interaction cross section ($\sigma_I$) measurements signal the possibility of a thick neutron surface. The large $\sigma_I$ of $^{23}$O is explained by $^{22}$O core + neutron in the 2$s_{1/2}$ orbital \cite{Kanungo2011}. This is consistent with its narrow one-neutron removal longitudinal momentum distribution \cite{CortinaGil2004,Sauvan2004} and its large Coulomb dissociation cross section \cite{Nociforo2005}. The neutron removal momentum distribution of $^{24}$O shows predominant valence neutron occupancy in the 2$s_{1/2}$ orbital \cite{Kanungo2009}. The matter radii derived from low-energy proton elastic scattering \cite{Lapoux2016} is systematically higher than from the $\sigma_I$ measurements. At energies below 100$A$ MeV medium modification effects of the nuclear interaction can lead to large uncertainty in the extraction of the radii. 

{\it {Ab initio}} calculations with chiral interactions as introduced in Ref.\cite{Lapoux2016} predicted the radii of oxygen isotopes. In-medium similarity renormalization group (IMSRG)  and Gorkov self-consistent Green's function theory (GGF) with the Entem-Machleidt (EM) chiral interaction resulted in smaller radii than with the NNLO$_{sat}$ interaction \cite{Lapoux2016}.  An increase of charge radius $\sim$ 0.03 - 0.05 fm is predicted between $^{16}$O and $^{24}$O  using the SRG evolved chiral interaction \cite{Binder2018}  and the $\Delta$-full interaction at N$^2$LO \cite{Hoppe2019}.  The NN+3N(lnl) chiral Hamiltonian and the NNLO$_{sat}$ interactions in the Gorkov self-consistent Green's function theory \cite{Soma2020} predicts a continuous increase of the charge radii with increasing mass number. 
In the relativistic mean field framework an ansatz simulating the pairing effect \cite{An2020} predicts charge radii with odd-even staggering. For neutron-rich isotopes they predict $^{20, 22}$O having a larger charge radius. There is no experimental information on the proton distribution radii beyond $^{18}$O. 

In this article, we present the first determination of root mean square point proton radii for $^{19-24}$O and those for the stable isotopes $^{16,18}$O derived from measurements of charge changing cross sections ($\sigma_{CC}$). The experiment was performed using the Fragment Separator FRS \cite{FRS} at GSI. The  $^{16-24}$O isotopes were produced by fragmentation of $^{40}$Ar accelerated to 1$A$ GeV which interacted with a Be target of thickness 6.3 g/cm$^2$. The fragments produced were separated and identified using the FRS by employing the event-by-event determination of mass to charge ratio ($A/Q$) and atomic number $Z$ information derived from the magnetic rigidity ($B\rho$), time-of-flight (TOF) and energy-loss ($\Delta E$). The isotopes were fully stripped hence $Q$ = $Z$. A schematic of the detector placement is shown in Fig1(a) and the particle identification is shown in Fig.1(b). The energy-loss of the fragments in a multi-sampling ionization chamber (MUSIC)~\cite{MUSIC} provided the $Z$ information.   The time of flight was measured between the dispersive mid-focal plane F2 and the achromatic final focal plane F4 using the fast plastic scintillators. Position sensitive Time Projection Chamber (TPC) detectors placed at these focal planes were used for beam tracking. The position information and the magnetic field provided the $B\rho$ determination of the incoming beam.  

\begin{figure}
\includegraphics[width=9cm, height=8cm]{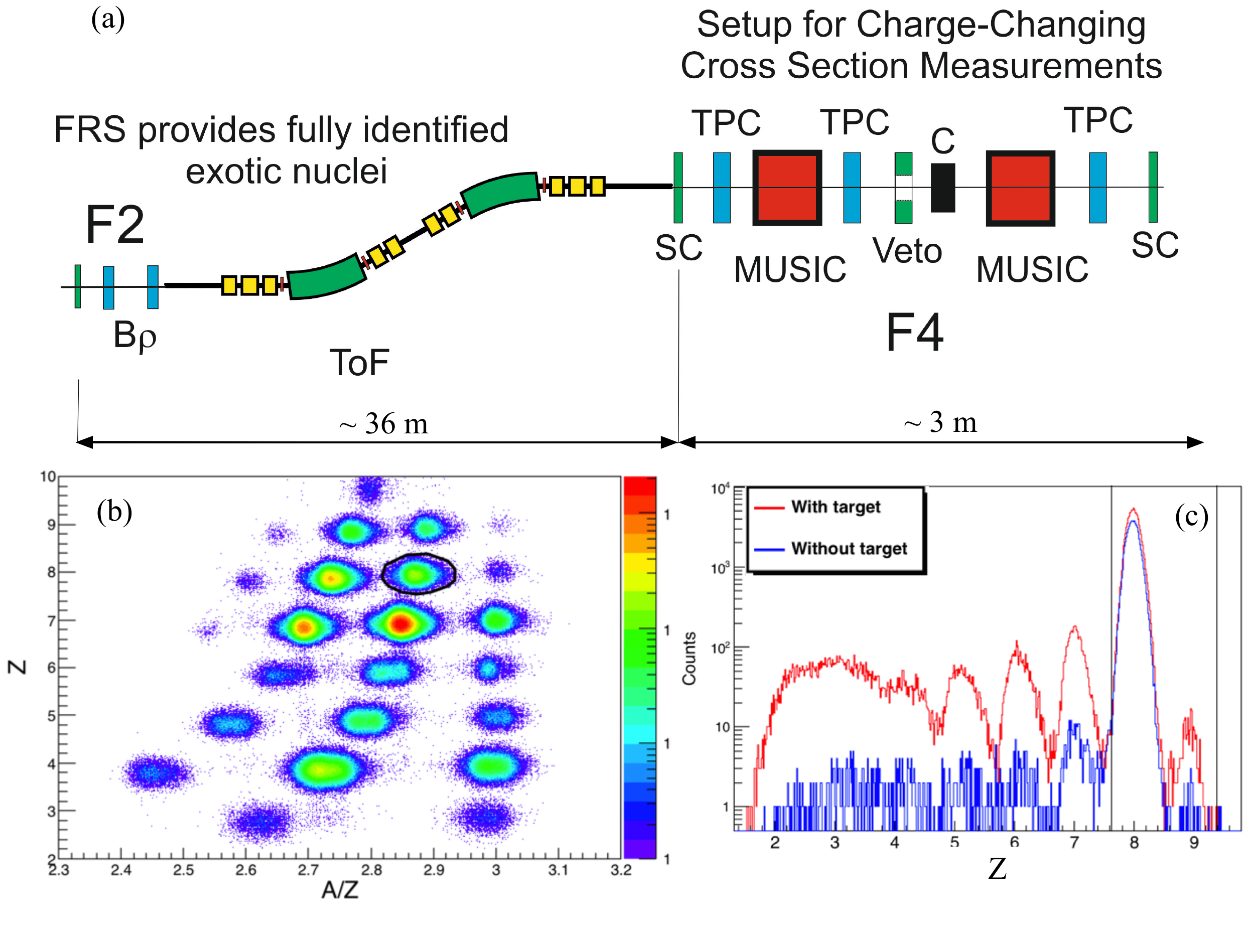}
\caption{\label{fig:epsart}  (a) Schematic view of the experiment setup at the FRS with detector arrangement at the final focus F4. (b) Particle identification before C target at F4. (c) $Z$ identification using MUSIC detector behind C target. The red / blue histogram shows data without / with C target. }
\end{figure}

The $\sigma_{CC}$ was measured with a 4.010 g/cm$^2$ thick C target placed at F4. The measurement was done using the transmission technique, where the ratio of the particles transmitted through the target without any loss of protons to the number of incoming particles gives the desired cross section for determining the root mean square radius of the point proton distribution, hereafter referred to as the proton radius.  For this measurement the number ($N_{in}$) of the incident nuclei $^{A}Z_{in}$ before the reaction target is identified and counted event-by-event. Behind the reaction target, the nuclei with charge $Z_{out} \geq Z_{in}$ are identified and counted on an event-by-event basis ($N_{Z \geq Z_{in}}$). The charge changing cross section is given by $\sigma_{CC}$ = $t^{-1}$ln(R$_{\rm{T}_{out}}$/R$_{\rm{T}_{in}}$). Here R$_{\rm{T}_{in}}$ and R$_{\rm{T}_{out}}$ are the ratios of $N_{Z \geq Z_{in}}$/$N_{in}$ with and without the reaction target, respectively and $t$ is the target thickness. Data without the reaction target was collected in order to account for losses due to interaction with the non-target materials. There is no uncertainty in $N_{in}$ due to freedom of any incident beam event selection in the event-by-event counting. 

In order to eliminate beam particle losses due to the restricted acceptance of the target and/or detectors the incident beam events were chosen with a restricted phase space.  This reduces the systematic uncertainty in the transmission ratio. A veto scintillator with a central aperture was placed in front of the target to reject beam events incident on the edges of the target scattered by matter upstream and multi-hit events that can cause erroneous reaction information in the MUSIC detector placed after the target. In the incident beam identification the estimated contamination from $Z$ = 7 and 9 are 6$\times$10$^{-5}$ and 2$\times$10$^{-5}$, respectively. 

In order to count the $^A$O beam events that did not undergo proton removal reactions in the target, the spectrum of the MUSIC detector placed after the target was used with the condition of the selected incoming $^A$O beam events for the $Z_{out} \geq 8$ identification (Fig.1c). The limits are chosen to be the 3.5$\sigma$ ends of the Z = 8 and 9 peaks. The $Z$ = 9 peak is included in the unreacted event counting because proton pickup or $(p,n)$ reactions leading to higher $Z$ do not involve interaction with protons in the projectile. Hence, for determining the proton radius these are unreacted proton events. The energy-loss in the TPC and plastic scintillator detectors placed further downstream of the target provided additional information to confirm the $Z$ identification as well as determine the detection efficiency of the MUSIC detector. The MUSIC detector resolution for $Z$ was $\sim$ 0.1 ($\sigma$). The estimated $Z_{out}$ = 7 contamination in the selection region of unreacted $Z_{out}$ is $\sim$ 5$\times$10$^{-5}$ which leads to an average uncertainty of $\pm$0.07 mb in the $\sigma_{CC}$. 

The measured cross sections and their one standard deviation total uncertainties are given in Table 1. This includes the target thickness uncertainty of $\sim$0.1\%. The systematic uncertainty from contaminants vary for the different isotopes ranging from 0.05 mb - 1 mb. The cross section for $^{16}$O aligns with the value 813(8) mb reported in Ref.\cite{Weber1990} at a slightly higher energy of 903$A$ MeV. The cross sections reported in Ref.\cite{Chulkov2000} at 930$\pm$44$A$ MeV are systematically higher as found also for other isotopic chains and have larger uncertainties making them unsuitable to accurately derive the proton radii.  

To extract the root mean square radii the measured $\sigma_{CC}$ are compared to cross sections calculated ($\sigma_{CC}^{cal}$) using the Glauber model framework \cite{SU16}. The formalism uses harmonic oscillator density profiles for the protons and neutrons in the projectile nucleus and the carbon target. The variation of the harmonic oscillator width yields projectile proton densities with different root mean square proton radii ($R_p$) which give different $\sigma_{CC}^{cal}$. The consistency of the measured $\sigma_{CC}$ and $\sigma_{CC}^{cal}$ determines the range of $R_p^{ex}$ that agrees with the data. The derived $R_p^{ex}$ are listed in Table 1. A good agreement of $R_p^{ex}$ and the root mean square point proton radii derived from electron scattering ($R_p^{(e^-)}$) is seen for $^{16,18}$O. This supports the successful determination of  $R_p^{ex}$ from the measured $\sigma_{CC}$. 
The gradual filling of neutrons in the  1$d_{5/2}$ orbital is found to decrease the  $R_p^{ex}$ progressively for $^{20 - 22,24}$O (Table 1 and Fig. 2 ). This is consistent with lower B(E2) values \cite{Raman2001}. A local minimum seen at $N$ = 14 is reflecting this new sub-shell closure. The consistent decrease in the proton radius for both $^{21}$N \cite{Bagchi2019} and $^{22}$O shows the $N$ = 14 sub-shell gap arises from the attractive $T$ = 0 monopole tensor interaction between the protons in the 1$p_{1/2}$ orbital and neutrons in the 1$d_{5/2}$ orbital. 

\begin{figure}
\includegraphics[width=6cm, height=12cm]{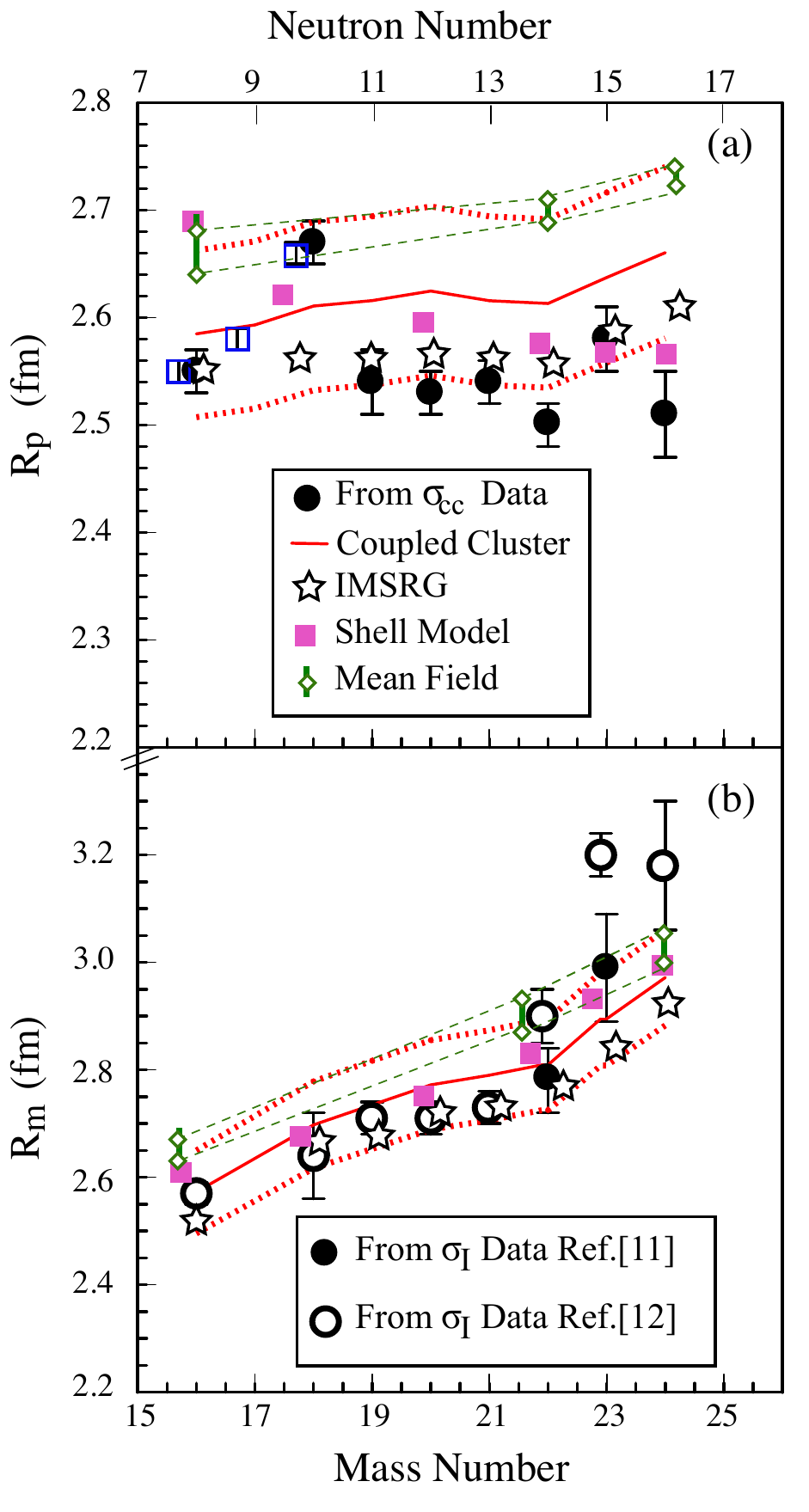}
\caption{\label{fig:epsart}  (a) $R_p^{ex}$ (filled circles), blue open squares show $R_p^{e^-}$. (b) $R_m^{ex}$ (Table 1) $\sigma_I$ from Ref.\cite{Ozawa2000} (open circles), Ref.\cite{Kanungo2011} (filled circles). The curves show predictions with coupled cluster theory for NNLO$_{sat}$ interaction (red curves). The dotted curves represent the $\pm$ 3$\%$ uncertainty of the theory. The predictions with NNLO$_{sat}$ interaction and the IMSRG model are shown by the star symbols. The pink squares show shell model predictions. The green bars and dashed lines show mean field results.}
\end{figure}

The proton radius of $^{23}$O increases due to its extended neutron density distribution where the valence neutron is occupying predominantly the 2$s_{1/2}$ orbital. The proton radius of $^{24}$O is found to be smaller than $^{23}$O but similar to that of $^{22}$O. This suggest the center-of-mass of the two valence neutrons in $^{24}$O is not greatly separated spatially from that of the core. The filling of the 2$s_{1/2}$ orbital also leads to stronger neutron binding of the two-valence neutrons in $^{24}$O due to pairing. 

Using the $R_p^{ex}$ determined in this work we find the point matter radius by analyzing the interaction cross sections ($\sigma_I$) reported in Refs.\cite{Ozawa2000,Kanungo2011}. At the high energies inelastic scattering cross section to bound excited states is negligible. Therefore, $\sigma_I$ = $\sigma_R$, the reaction cross section. The nucleon-target profile function in the Glauber model (NTG) \cite{Ibrahim2000}  with the profile function given in Ref.\cite{Ibrahim2008} is used for calculating $\sigma_R^{cal}$, for which harmonic oscillator densities of protons and neutrons for $^A$O are adopted. The densities that result in $\sigma_R^{cal}$ agreeing with the measured $\sigma_I$ yield the point matter radii ($R_m^{ex}$) that are listed in Table 1. The $R_m^{ex}$ of $^{19-22}$O shows a small gradual increase, trend that is broken at $^{23}$O which shows a larger increase in $R_m^{ex}$. We note that the later measurement of interaction cross section of $^{22,23}$O \cite{Kanungo2011} yield matter radii that agree with the description of $^{23}$O in a $^{22}$O core plus neutron model with large spectroscopic factor for the neutron in the 2$s_{1/2}$ orbital. This is consistent with the observations from knockout reactions \cite{CortinaGil2004} and Coulomb dissociation \cite{Nociforo2005}.

\begin{table}
\caption{\label{tab:table1} Secondary beam energies at the entrance of the C target, measured $\sigma_{cc}$ and the root mean square proton and matter radii derived from the data for the oxygen isotopes.}
\begin{ruledtabular}
\begin{tabular}{llllll}
Isotope & E/A &$\sigma_{cc}^{ex}$&$R_p^{ex}$&$R_p^{(e^-)}$ &$R_m^{ex}$\\
&(MeV)&(mb)&(fm)&(fm)&(fm)\\
\hline
$^{16}$O& 857  & 848(4)&2.54(2) &2.55(1) &2.57(2) \\
$^{18}$O& 872 & 879(5)&2.67(2)&2.66(1) &2.64(8)\\
$^{19}$O& 956 & 852(7)&2.55(3)&  &2.71(3)\\
$^{20}$O& 880 & 846(4)&2.53(2)&  &2.71(3)\\
$^{21}$O& 937 & 847(6)&2.53(2)&  &2.73(3)\\
$^{22}$O& 937 & 837(3)&2.50(2)&  &2.78(6)\footnote{\label{note1} $\sigma_I$ \cite{Kanungo2011}}\\
$^{22}$O&   &  & &  & 2.90(5)\footnote{\label{note2} $\sigma_I$ \cite{Ozawa2000}}\\
$^{23}$O& 871 & 857(8)&2.58(3)&  &2.99(10)$^a$\\
$^{23}$O&   &  & &  &3.20(4)$^b$\\
$^{24}$O& 866 & 839(11)&2.51(4)&  &3.18(12)\\
\end{tabular}
\end{ruledtabular}
\end{table}

{\it Ab-initio} coupled-cluster and valence-space (VS) IMSRG computations are performed employing the chiral NN+3N interaction NNLO$_{sat}$~\cite{ekstrom2015}, which generally reproduces absolute and relative trends in radii across isotopic chains in both the $sd$~\cite{Heyl21Al} and $pf$ shells~\cite{Groo20Cu,Malb22NiRch}. 
For the coupled-cluster calculations we employ the singles-and-doubles (CCSD) approximation~\cite{bartlett2007}, and start from an axially deformed Hartree-Fock reference state (assuming a prolate shape) following Refs.~\cite{novario2020,koszorus2021}. 
In the VS-IMSRG, an approximate unitary transformation is constructed to decouple a core and effective valence-space Hamiltonian~\cite{Stro19ARNPS,Stro17ENO,Miya22Heavy} diagonalized with the KSHELL code~\cite{KSHELL}. Applying the same transformation to the point proton radius operator we further construct an effective valence-space operator consistent with the Hamiltonian. 
Other details of the {\it ab initio} radii calculations can be found in Ref.~\cite{Malb22NiRch}.
Combining the effects from neglected many-body correlations, model-space truncations, and
symmetry breaking we estimate an uncertainty of $\pm 3\%$ on the coupled-cluster computations which is correlated for the point nucleon radii, hence negligible for relative quantities. 

The $R_p^{ex}$ are compared (Fig. 2a) with the CCSD predictions (red curves). The NNLO$_{sat}$ interaction reproduces binding energies of oxygen isotopes~\cite{ekstrom2013, ekstrom2015}. It reproduces also the trends of $R_p^{ex}$ reasonably well for neutron-rich isotopes predicting a radius dip at $N$ = 14 consistent with the data. The IMSRG (star symbols) results with the NNLO$_{sat}$ interaction from Ref.\cite{Lapoux2016},  are within the uncertainty band of the CCSD results and also show a local minimum at $N$ = 14. In contrast, Ref.\cite{An2020} predicts an increase in the charge radius of $^{22}$O. The NN+3N(lnl) chiral interaction predictions \cite{Soma2020} are smaller than the data showing an improved description of the nuclear interaction by NNLO$_{sat}$. Within the data uncertainties the $R_p^{ex}$ of the doubly closed shell nuclei $^{16}$O and $^{24}$O are similar, with an indication, from the central values, of a possible reduction in  $^{24}$O due to its stronger proton binding. 

Shell model calculations with the YSOX Hamiltonian  \cite{Yuan2012} are shown in Fig. 2 and 3 (filled squares).
Occupation numbers for the orbits obtained with the YSOX are used to evaluate the proton and matter radii as well as the neutron skin thickness. 
The proton orbits are obtained in a Woods-Saxon potential with the standard parameters \cite{BM}, the resulting proton radii (Fig. 2a filled pink squares) are in fair agreement with the data.
Using these proton radii, the matter radii are computed using harmonic oscillator functions for neutron orbits with $\hbar\omega$ = 45/$A^{1/3}$ -25/$A^{2/3}$ except for the 2$s_{1/2}$ orbital in $^{23,24}$O,
which are obtained by the three-body model with an inert $^{22}$O core plus 2$s_{1/2}$ valence neutron description \cite{TSuzuki2016}. 
Use is made of a low-energy limit of the valence neutron-neutron interaction, which can reproduce the known $NN$ scattering length and effective range \cite{TSuzuki2016}.

Mean-field Hartree-Fock calculations with Sk3, SLy4 and  SKM forces (Fig. 2 green bars and dashed lines) show proton radii of $^{22}$O and $^{24}$O larger than that of $^{16}$O, independent of the Skyrme forces. This is contrary to the data trend. Inclusion of the coupling to the monopole resonances and improvements of proton-neutron interaction need to be considered in future for the study of neutron-rich nuclei in the mean-field approximation.
 
The point matter radii are compared to the model predictions in Fig. 2b. The coupled cluster theory predictions with $\pm$ 3$\%$ uncertainty band is shown by the red solid and dotted curves, respectively. The IMSRG calculations performed in this work are shown by the star symbols.  An overall good agreement with the data affirms the NNLO$_{sat}$ interaction to be successful in predicting the radii from stable isotopes to the drip-line. The shell model predictions also agree well with the data (Fig. 2b filled pink squares). The mean field model predictions are shown by the green bars and dashed lines. The predicted radii align better with the data for $^{22,23}$O from Ref.\cite{Kanungo2011} which are also consistent for a core + neutron(2$s_{1/2}$) model for $^{23}$O. We therefore, use these data to derive the neutron skin thickness shown in Fig.3. The neutron skin thickness is found by  $R_n - R_p$ = $\sqrt{(A/N)R_m^2 - (Z/N)R_p^2} - R_p$, where $A, N, Z$ are the mass number, neutron number and proton number, respectively. $R_m$, $R_n$ and $R_p$ are the point matter, point neutron and point proton $rms$ radii, respectively. The data reveal a thick neutron surface for $^{22-24}$O. The skin thickness predicted by the coupled cluster model and the IMSRG are similar but underestimate the data beyond $^{21}$O. The shell model predictions of neutron skin thickness (Fig.3  filled squares) successfully describe the large neutron surface for $^{22,23}$O.

\begin{figure}
\includegraphics[width=6cm, height=6cm]{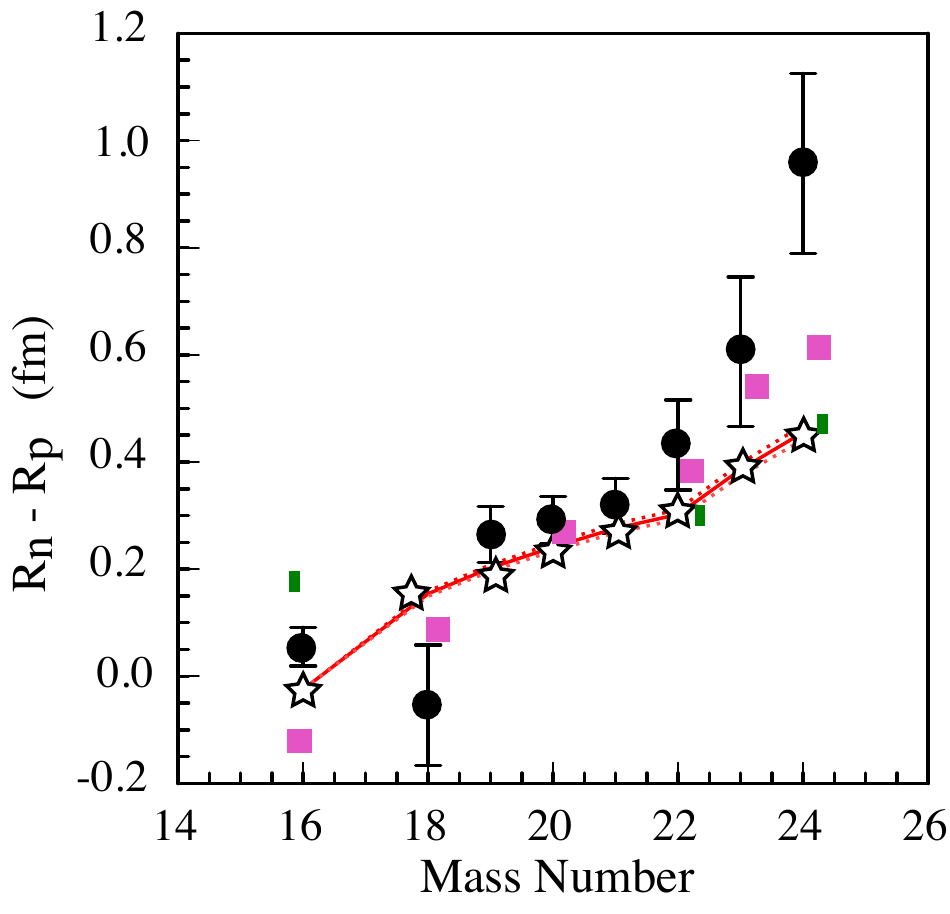}
\caption{\label{fig:epsart}  Neutron skin thickness data from $R^{ex}_p$ and $R^{ex}_m$  filled circles. The red solid (dotted) curves represent the predictions ($\pm$ 3$\%$ uncertainty) from coupled cluster theory using  the NNLO$_{sat}$ interaction. The star symbols represent predictions from the IMSRG calculations. The pink filled squares are shell model results. Green bars show mean-field results.}
\end{figure}

In summary, the point proton radii of $^{16,18-24}$O derived from measurements of charge changing cross sections show an extended radius for $^{23}$O and local minimum for $^{22}$O that relates to the $N$ = 14 sub-shell closure due to the tensor force. The doubly magic nature of $^{22,24}$O does not hinder neutron skin development, which rapidly increases for $^{22-24}$O. Shell model predictions reproduce the observed neutron skin. {\it Ab initio} predictions with the NNLO$_{sat}$ chiral interaction agree within theoretical uncertainty with $R_p^{ex}$ showing a dip for $^{22}$O as observed in the data. The predictions for neutron skin thickness of $^{22-24}$O underestimate the data. The data therefore open new avenues for refining the chiral interaction.

The authors are thankful for the support of the GSI accelerator staff
and the FRS technical staff for an efficient running of the
experiment and  S.R. Stroberg for the imsrg++ code~\cite{Stro17imsrg++} used to perform these calculations.. The support from NSERC, Canada for this work is gratefully
acknowledged. The support of the PR China government and Beihang university under the
Thousand Talent program is gratefully acknowledged.  The experiment is
partly supported by the grant-in-aid program of the Japanese
government under the contract number 23224008. 
G.~H was supported by the Office of Nuclear Physics, U.S. Department of Energy, under
grant DE-SC0018223 (NUCLEI SciDAC-4 collaboration), and
contract No. DE-AC05-00OR22725 with UT-Battelle, LLC (Oak Ridge National Laboratory)
Computer time was provided
by the Innovative and Novel Computational Impact on Theory and
Experiment (INCITE) programme and an allocation of computing resources on Cedar at WestGrid and Compute Canada. This research used resources of the Oak
Ridge Leadership Computing Facility located at Oak Ridge National
Laboratory, which is supported by the Office of Science of the
Department of Energy under contract No. DE-AC05-00OR22725. 
This work was supported in part by JSPS KAKENHI grant No. JP18K03635, JP19K03855, JP19H05145 and JP21H00117. This work was supported in part by the Deutsche Forschungsgemeinschaft (DFG, German Research Foundation) -- Project-ID 279384907 -- SFB 1245. R.K. thanks L. Geng for providing the numerical values of the radii in Ref.\cite{An2020}.



\end{document}